\documentclass{aa}
\usepackage{times,graphicx}
\usepackage{supertabular}
\usepackage{curves,epic,eepic}
\usepackage{float}

\begin{document}

\thesaurus{03.13.5; 
           13.07.1;} 
            \title{Optical and near-infrared observations of the GRB 970616
                error box~\thanks{Based on observations
                collected at the German-Spanish Astronomical Centre, Calar
                Alto, operated by the Max-Planck-Institut f\"ur Astronomie,
                Heidelberg, jointly with the Spanish National Commission
                for Astronomy.}$^{,}$\thanks{Based on observations carried
                out at the Danish 1.54-m Telescope on the European Southern
                Observatory, La Silla, Chile}}

   \author{J. Gorosabel
          \inst{1}
   \and A. J. Castro-Tirado
          \inst{1,2}
   \and H. Pedersen
          \inst{3}
   \and J. Greiner
          \inst{4}
   \and D. Thompson
          \inst{5}
   \and M. Guerrero 
          \inst{6}
   \and A. Oscoz
          \inst{6}
   \and N. Sabalisck
          \inst{6}
   \and E. Villaver
          \inst{6}
   \and N. Lund
          \inst{7}}

   \offprints{J. Gorosabel (jgu@laeff.esa.es)}

   \institute{Laboratorio de Astrof\'{\i}sica Espacial y F\'{\i}sica
     Fundamental (LAEFF-INTA), P.O. Box 50727, E-28080 Madrid, Spain.
\and Instituto de Astrof\'{\i}sica de Andaluc\'{\i}a (IAA-CSIC), P.O. Box
     03004, E-18080 Granada, Spain.
\and Copenhagen University  Observatory, Juliane Maries Vej 30 DK 2100 
     Copenhagen $\O$, Denmark.
\and Astrophysikalisches Institut Potsdam, D-14482 Potsdam, Germany.
\and Max-Planck-Institut f\"ur Astronomie, Heidelberg, Germany.  
\and Instituto de Astrof\'{\i}sica de Canarias, E-38200 La Laguna,
     Tenerife, Spain.
\and  Danish Space Research Institute, Juliane Maries Vej 30 DK-2100 
   Copenhagen $\O$, Denmark.}
         
   \date{Received date; accepted date}
   
   \titlerunning{Optical/IR study of the GRB 970616 field.}


   \maketitle

   \begin{abstract} 
     We report on near-infrared and optical observations of the GRB 970616
     error box and of the X-ray sources discovered by ASCA and ROSAT in the
     region. No optical transient was found either within the IPN band or
     in the X-ray error boxes, similarly to other bursts, and we suggest
     that either considerable intrinsic absorption was present (like GRB
     970828) or that the optical transient displayed a very fast decline
     (like GRB 980326 and GRB 980519).

\keywords{Methods: observational, Gamma rays: bursts}
   \end{abstract}

%

\section{Introduction}

The error box of the BATSE gamma-ray burst GRB 970616 (Connaughton et al.
1997) was scanned by RXTE four hours after the event (Marshall et al.
1997). Observations revealed a previously unknown X-ray source inside the
error box whose position was consistent with the one provided later on by
the Interplanetary Network (IPN), yielding a combined RXTE/IPN trapezoidal
error box (Hurley et al. 1997). Four days later, the ASCA satellite
observed the RXTE/IPN error box and detected four X-ray sources.  One of
them, A\#1 according to the nomenclature of Murakami et al. (1997), was
suggested as the X-ray counterpart to GRB 970616. Two candidates were
initially proposed by Galama et al. (1997) and Udalski (1997), but other
observers (Pahre et al. 1997, Dey et al. 1997, Wheeler et al. 1997,
Castro-Tirado et al. 1997) failed to confirm these.  Observations performed
by ROSAT during June 23-25 revealed eleven faint X-ray sources, lying three
of them within the RXTE/IPN error box.  One of the three sources, R\#2,
following Greiner et al. (1997), was consistent with the ASCA variable
source A\#1, showing a flux five times lower than the one measured by ASCA.
The 3$\sigma$ upper limit to the unabsorbed ROSAT flux of the source A\#4,
implies a factor of seven lower than the previously measured by ASCA.  The
fading observed during the ASCA observation could suggest that A\#1 is the
X-ray counterpart to the GRB.  However, the presence of another decaying
X-ray source inside the RXTE/IPN error box keeps open the association with
the GRB. A massive cluster of galaxies $2^{\prime}$ away from the RXTE/IPN
error box is reported by Ben\'{\i}tez et al.  1999. Fig 1 shows the
locations of the cluster and of the candidates proposed by Galama et al.
(1997) and Udalski (1997).


\section{Observations}
\label{970616observations}
\subsection{Near-IR observations}

Observations in   the   K$^{\prime}$-band were performed  with   the  3.5-m
telescope (+OMEGA) of the Calar Alto Observatory on  1997 June 21.18, 25.17
and 26.18 UT.   Also J-band images were   taken  with the  2.2-m  Telescope
(+MAGIC) on June 25.14. The frames  were centered on the candidate proposed
by  Galama et al.  (1997).   Additional K$^{\prime}$-band observations were
performed on 1998 Oct. 5.125 with the 3.5-m telescope.

\subsection{Optical Observations}
Optical observations were mainly  taken at La  Silla with the Danish 1.54-m
telescope equipped with  DFOSC, whose field  of view ($13^{\prime}.6 \times
13^{\prime}.6$)  enabled  us to  cover the  four  ASCA sources   (see Fig.  
1).~Table 1 displays the observing log.

\begin{table}[H]
\begin{center}
\caption{\label{970616table2} \small Observing log of the optical observations 
  performed with the 1.54-m Danish telescope.}
  \begin{tabular}{lccccc}
  \hline 
\scriptsize Date & \multicolumn{4}{c}{\scriptsize Exposure time (10$^3$ sec)} \\ 
\scriptsize of 1997 & \scriptsize B &\scriptsize  V &\scriptsize  R &\scriptsize  i &\scriptsize free\\ 
   \hline 
\scriptsize     June 25 & -    & -    &\scriptsize  8.9   &  -   &   - \\ 
\scriptsize     June 26 & -    & -    & -     &\scriptsize  7.8  &   - \\ 
\scriptsize     June 27 & -    & -    &\scriptsize  9.6   &   -  &   - \\ 
\scriptsize     June 28 & -    & -    &\scriptsize  10.2  &   -  &   - \\ 
\scriptsize     June 29 & -    & -    &\scriptsize  11.2  &   -  &   - \\
\scriptsize     June 30 & -    &\scriptsize  2.0  &\scriptsize  9.0   &   -  &   - \\
\scriptsize     July  1 & -    &\scriptsize  1.2  &\scriptsize  10.5  &   -  &   - \\
\scriptsize     July  2 &\scriptsize  3.6  & -    &\scriptsize  7.2   &   -  &   - \\
\scriptsize     July  3 &\scriptsize  1.2  & -    & -     &   -  &   - \\
\scriptsize     July  4 &\scriptsize  1.2  & -    & -     &   -  &   - \\
\scriptsize     July  6 & -    & -    &\scriptsize  3.6   &   -  & \scriptsize 0.9 \\
\scriptsize     July  7 & -    & -    & -     &   -  &\scriptsize  5.6 \\
   \hline
\scriptsize      Total 26.02$^h$=&\scriptsize  6.0&\scriptsize +3.2&\scriptsize +70.2&\scriptsize +7.8&\scriptsize +6.5\\
 \\ 
\end{tabular} 
\end{center}
\end{table}

Additional R-band imaging was obtained on 1997 July 20.199, 20.207 and
20.257 with the 0.8-m IAC80 telescope at the Observatorio del Teide. A 1020
$\times$ 1024 pixels CCD provides a $7^{\prime}.4 \times 7^{\prime}.4$
field of view. Limiting magnitude was R $\sim$ 20 on the 360-s exposures.

\section{Discussion and conclusion}

Optical and IR images of the RXTE error box were obtained starting 106
hours after the event.  
No variable optical counterpart was found within the A\#1 and A\#4 X-ray
source error circles in the RXTE\-/IPN annulus, being the variation
$\Delta$B $\leq$ 0.3, $\Delta$V $\leq$ 0.2, $\Delta$R $\leq$ 0.2 and
$\Delta$i $\leq$ 0.3 for optical objects with B $\leq$ 23, V $\leq$ 23, R
$\leq$ 24 and i $\leq$ 22. In fact, no objects brighter than R = 24.2 are
seen within the error box of the ROSAT R\#2 source detected at the ASCA
position, consistently with the magnitudes reported by Groot et al.  (1997).
Therefore, either there is a considerable intrinsic absorption (as in the
case of GRB 970828, Yoshida et al. 1999) or the transient optical emission
displayed a fast decline, as seen in GRB 980326 (Groot et al.  1998) and
GRB 980519 (Hjorth et al. 1999).

No IR variability has been found in a $5^{\prime} \times 4^{\prime}$ region
centered at the position of the candidate proposed by Galama et al. (1997).
This object was observed at J = 18.3 $\pm$ 0.2, and K$^{\prime}$ = 17.4
$\pm$ 0.3 and did not show any obvious IR/optical variation in the
K$^{\prime}$-band. The color is consistent with this of a late-type star,
as suggested by Pahre et al.  (1998) and it seems to be a non-variable star
(Groot et al. 1997). Any fading or increase was $\leq$ 0.5 mag for
K$^{\prime} \leq$ 17.0.

Concerning the above mentioned cluster of galaxies  (the long arrow in Fig. 
1), it is probably  related to the X-ray source  A\#2, which is outside the
IPN band, and consequently not related to the GRB.

 \begin{figure}[t]
   \centering
   \resizebox{\hsize}{!}{\includegraphics{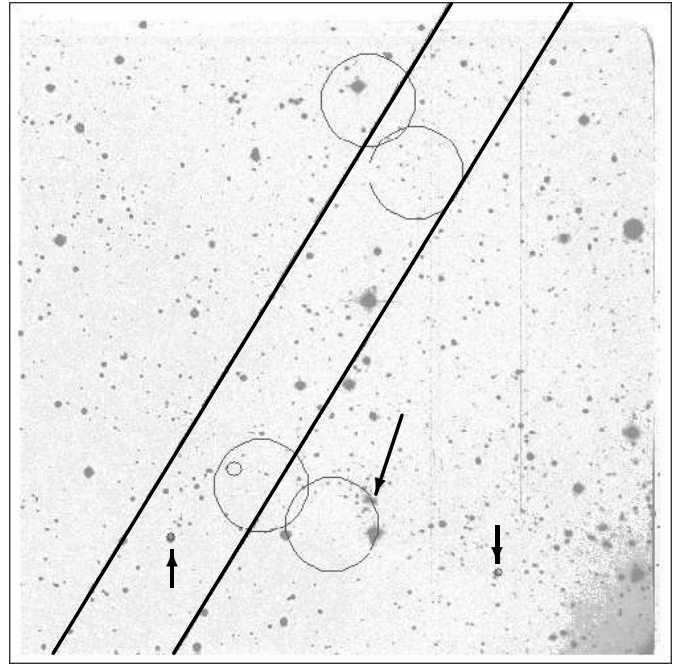}}
\allinethickness{0.5mm}
\put(-102,94){\vector(-1,-3){10}}
\put(-189,29){\vector(0,1){14}}
\put(-66,52){\vector(0,-1){14}}
\drawline[5](-234,4)(-83.5,250)
\drawline[5](-188.5,4)(-38,250)

    \caption[R-band image of the field of GRB 970616]
    {A coadded R-band image of the field of GRB 970616 obtained with the
      Danish 1.54-m telescope. The total exposure time is 19.5 hours and
      the limiting magnitude  R = 24.2. The straight thick lines represent the
      IPN arc determined by Ulysses and BATSE. The two circles fully
      included in the IPN band are A\#4 (large, at the top) and R\#2 (small
      at the bottom), whereas the circles partially included are A\#3(top)
      and A\#1 (bottom). The circle totally outside the IPN band is A\#2.
      R\#1 and R\#3 are outside the field of view.  The two short arrows
      show the variable objects proposed by Galama et al. (1997) -inside 
      the IPN band- and Udalski (1997) -outside the IPN band-.  The field of
      view is $13^{\prime}.6 \times 13^{\prime}.6$ and is totally within the
      RXTE/ASM error box. North is to the top and
      east to the left. The long arrow indicates the massive cluster of
      galaxies found by Ben\'{\i}tez et al.  (1999).}
    \label{970616figure2}
 \end{figure}  

\begin{acknowledgements}
  We whish to thank A. Yoshida for fruitful conversations and the referee
  N. Masetti for useful suggestions.  This work has been partially
  supported by Spanish CICYT grant ESP95-0389-C02-02.
\end{acknowledgements}


\bibliographystyle{aa} 

\begin{thebibliography}{}

\bibitem[\protect\citeauthoryear{{Ben\'{\i}tez} et~al.}{{Ben\'{\i}tez}
  et~al.}{1998}]{Beni98}
{Ben\'{\i}tez} N., et al. 1999, in preparation.

\bibitem[\protect\citeauthoryear{{Castro-Tirado} et~al.}{{Castro-Tirado}
  et~al.}{1997}]{Cast97d}
{Castro-Tirado} A.~J., et al.  1997, IAU Circ., 6688.

\bibitem[\protect\citeauthoryear{{Connaughton} et~al.}{{Connaughton}
  et~al.}{1997}]{Conn97}
{Connaughton} V.,  et al. 1997, IAU Circ., 6683.

\bibitem[\protect\citeauthoryear{{Dey}, {Stern}, \& {Bunker}}{{Dey}
  et~al.}{1997}]{Dey97}
{Dey} A., et al.  1997, IAU Circ., 6696

\bibitem[\protect\citeauthoryear{{Galama} et~al.}{{Galama}
  et~al.}{1997}]{Gala97e}
{Galama} T., et al. 1997, IAU Circ.,  6687.

\bibitem[\protect\citeauthoryear{{Greiner} et~al.}{{Greiner} et~al.}{1997}]{Grein97b}
{Greiner} J., et~al. 1997, IAU Circ., 6722.

\bibitem[\protect\citeauthoryear{{Groot} et~al.}{{Groot} et~al.}{1997}]{Groo97}
{Groot} P. J., et~al. 1997, IAU Circ., 6723. 

\bibitem[\protect\citeauthoryear{{Groot} et~al.}{{Groot} et~al.}{1998}]{Groo98}
{Groot} P. J., et~al. 1998, ApJ, 502, L123.

\bibitem[\protect\citeauthoryear{{Hjorth} et~al.}{{Hjorth} et~al.}{1998}]
{Hjor98}{Hjorth} J., et al.  1999,  A\&AS, this issue.

\bibitem[\protect\citeauthoryear{{Hurley}, {Kouveliotou}, \&
  {Marshall}}{{Hurley} et~al.}{1997}]{Hurl97g}
{Hurley} K., et al. IAU Circ., 6687.

\bibitem[\protect\citeauthoryear{{Marshall} et~al.}{{Marshall}
  et~al.}{1997}]{Marsh97b}
{Marshall} F.~E., et al. 1997,  IAU Circ., 6683.

\bibitem[\protect\citeauthoryear{{Murakami} et~al.}{{Murakami}
  et~al.}{1997}]{Mura97a}
{Murakami} T., et~al. IAU Circ., 6687.

\bibitem[\protect\citeauthoryear{{Pahre} et~al.}{{Pahre}
  et~al.}{1997}]{Pahre97}
{Pahre} M.~A., et~al. 1997, IAU Circ., 6691.

\bibitem[\protect\citeauthoryear{{Udalski}}{{Udalski}}{1997}]{Udalski97}
{Udalski} A. 1997, IAU Circ., 6690.

\bibitem[\protect\citeauthoryear{{Wheeler} et~al.}{{Wheeler}
  et~al.}{1997}]{Wheeler97}
{Wheeler} J.~C., et~al. 1997, IAU Circ., 6697.

\bibitem[\protect\citeauthoryear{{Yoshida} et~al.}{{Yoshida} et~al.}{1998}]
{Yosh98}
{Yoshida} A. et~al. 1999, A\&AS, this issue.

\end{thebibliography}

\end{document}